\documentclass[preprint,12pt,aps,amsmath,amssymb,nofootinbib,superscriptaddress,showkeys,showpacs]{revtex4}
\usepackage{epsfig} 
\usepackage{amsmath} 
\usepackage{graphicx} 
\def\beq{\begin{equation}}
\def\eeq{\end{equation}}
\def\bea{\begin{eqnarray}}
\def\eea{\end{eqnarray}}
\def\nn{\nonumber}

\def\dub{\delta_{ub}}
\def\dubp{\delta_{ub'}}
\def\dcbp{\delta_{cb'}}
\def\v44{V_{4 \times 4}}
\def\v34{V_{3 \times 4}}

\def\V{\widetilde{V}}

\def\lesssim{\mathrel{\hbox{\rlap{\hbox{\lower4pt\hbox{$\sim$}}}\hbox{$<$}}}} 
\def\gtrsim{\mathrel{\hbox{\rlap{\hbox{\lower4pt\hbox{$\sim$}}}\hbox{$>$}}}} 
\def\bra {\langle}
\def\ket {\rangle}
\def\bs{B_s^0}
\def\bsbar{{\bar B}_s^0}
\def\roughly#1{\mathrel{\raise.3ex\hbox
{$#1$\kern-.75em\lower1ex\hbox{$\sim$}}}}

\begin{document}  

\title{Constraints on the Four-Generation Quark Mixing Matrix 
from a Fit to Flavor-Physics Data}

\author{Ashutosh Kumar Alok}
\email{alok@lps.umontreal.ca}
\affiliation{Physique des Particules, Universit\'e de
Montr\'eal,\\ C.P. 6128, succ.\ centre-ville, Montr\'eal, QC,
Canada H3C 3J7}

\author{Amol Dighe}
\email{amol@theory.tifr.res.in}
\affiliation{Tata Institute of Fundamental Research, Homi
Bhabha Road, Mumbai 400005, India}
     
\author{David London}
\email{london@lps.umontreal.ca}
\affiliation{Physique des Particules, Universit\'e de
Montr\'eal,\\ C.P. 6128, succ.\ centre-ville, Montr\'eal, QC,
Canada H3C 3J7}

\date{\today} 

\preprint{UdeM-GPP-TH-10-188}
\preprint{TIFR/TH/10-32}


\begin{abstract}
In the scenario with four quark generations, we perform a fit using
flavor-physics data and determine the allowed values -- preferred
central values and errors -- of all of the elements of the $4 \times
4$ quark mixing matrix.  In addition to the direct measurements of
some of the elements, we include in the fit the present measurements
of several flavor-changing observables in the $K$ and $B$ systems that
have small hadronic uncertainties, and also consider the constraints
from the vertex corrections to $Z\to b\bar{b}$.  The values taken for
the masses of the fourth-generation quarks are consistent with the
measurements of the oblique parameters and perturbativity of the
Yukawa couplings.  We find that $|\V_{tb}| \ge 0.98$ at $3\sigma$, so
that a fourth generation cannot account for any large deviation of
$|V_{tb}|$ from unity. The fit also indicates that all the new-physics
parameters are consistent with zero, and the mixing of the fourth
generation with the other three is constrained to be small: we
obtain $|\V_{ub'}|< 0.06$, $|\V_{cb'}|< 0.027$, and $|\V_{tb'}|< 0.31$
at $3\sigma$.  Still, this does allow for the possibility of
new-physics signals in $B_d$, $B_s$ and rare $K$ decays.
\end{abstract}

\maketitle 

\newpage

\section{Introduction}

There is no unequivocal theoretical argument which restricts the
number of quark generations to three as in the standard model (SM).
An additional fourth generation (SM4) is one of the simplest
extensions of the SM, and retains all of its essential features: it
obeys all the SM symmetries and does not introduce any new ones.  At
the same time, it can give rise to many new effects, some of which may
be observable even at the current experiments \cite{Holdom:2009rf}.
Even though the fourth-generation quarks may be too heavy to have been
produced at the pre-LHC colliders, they may still affect low-energy
measurements through their mixing with the lighter quarks.  The
up-type quark $t'$ would contribute to $b \to s$ and $b \to d$
transitions at the 1-loop level, while the down-type quark $b'$ would
contribute similarly to $c \to u$ and $t \to c$.

The addition of a fourth generation to the SM leads to a $4 \times 4$
quark mixing matrix CKM4, which is an extension of the
Cabibbo-Kobayashi-Maskawa (CKM) quark mixing matrix in the SM.  The
parametrization of this unitary matrix requires six real parameters
and three phases.  The additional phases can lead to increased CP
violation, and can provide a natural explanation for the deviations
from the SM predictions seen in some measurements of CP violation in
the $B$-meson system
\cite{Hou:2005yb,Hou:2006mx,Soni:2008bc,Soni:2010xh,Buras:2010pi,Hou:2010mm}.
A heavy fourth generation can play a crucial role in the dynamical
generation of the electroweak (EW) symmetry breaking
\cite{EW-breaking}.  Also, the large Yukawa couplings of the fourth
generation quarks, together with the possible large phases, can help
efficient EW baryogenesis \cite{EW-transition}.

The EW precision measurements of the oblique parameters $S$ and $T$
imply strong correlations between the masses of the fourth-generation
quarks \cite{EW-constraint, Kribs:2007nz}. The parameter space of
fourth-generation masses with minimal contributions to $S$ and $T$,
and in agreement with all experimental constraints, is
\cite{Kribs:2007nz,erler-langacker}
\bea 
m_{t'} & \ge & 400~{\rm GeV}~, \nn \\
m_{t'}-m_{b'} &\simeq& \left(1+\frac{1}{5} \,\frac{m_H}{115\,
{\rm GeV}}\right) \times {50~ \rm GeV} ~,
\label{EWconstraints}
\eea 
where $m_{t'}, m_{b'}$ and $m_H$ are the masses of $t'$, $b'$, and the
Higgs boson $H$, respectively.  On the other hand, the perturbativity
of the Yukawa coupling implies that $m_{t'} \lesssim \sqrt{2\pi}
\langle v \rangle \approx 600$ GeV, where $\langle v \rangle$ is the
vacuum expectation value of the Higgs. Arguments based on the
unitarity of partial S-wave scattering amplitudes for color-singlet,
elastic, same-helicity $t'$-$\bar{t'}$ scattering at tree level
restrict $m_{t'} \lesssim \sqrt{4\pi/3} \langle v \rangle \approx 500$
GeV \cite{Chanowitz:1978mv, Marciano:1989ns}.  Thus, the
fourth-generation quark masses are constrained to a narrow band, which
increases the predictivity of the SM4.

The quark-mass bounds above may be somewhat relaxed with the
introduction of heavy fourth-generation leptons, which help in
partially cancelling out the effect of the fourth generation 
on the $S$ and $T$ parameters.
Even in the absence of any quark-lepton cancellation,
the EW precision measurements restrict 
\cite{Kribs:2007nz,erler-langacker}
\bea
m_{l'}-m_{\nu'} &\simeq& (30{\hbox{-}}60)~ {\rm GeV} ~, \nn
\eea
where $m_{l'}$ and $m_{\nu'}$ are the masses of the fourth-generation
charged lepton $l'$ and neutrino $\nu'$, respectively.  Thus even in
the absence of any fine-tuned cancellations, there is a significant
allowed range for the masses of the fourth-generation fermions, which
is, in fact, not beyond the reach of the LHC.  The invisible decay
width of the $Z$ boson constrains the mass of the fourth-generation
neutrino to be greater than 45 GeV.  Though one would need a special
mechanism leading to a massive fourth-generation neutrino and three
ultralight SM neutrinos, phenomenologically this is perfectly allowed.

In order to make concrete SM4 predictions, the first step is to
determine the elements of CKM4.  This involves not only fixing the
values of the new parameters, but also re-evaluating those of the SM.
This is because not all elements of the CKM matrix are measured
directly. For example, the bounds on $|V_{td}|$ and $|V_{ts}|$ are
obtained from decays involving loops, and these are rather weak. And
though $|V_{tb}|$ is measured in the tree-level decay $t \to bW$, its
value is not that precise: the direct measurement at the Tevatron from
single top production gives $|V_{tb}| = 0.88 \pm 0.07$
\cite{Abazov:2009ii,Aaltonen:2009jj,Group:2009qk}. Now, $|V_{tb}| = 1$
is predicted in the SM to an accuracy of $10^{-3}$.  Although the
Tevatron value is consistent with the SM prediction, it can also be as
small as $0.67$ at $3\sigma$. Thus, the values of the elements
$V_{tq}$ ($q = d,s,b$) are not obtained through measurements. Rather,
they are mainly determined using the unitarity of the $3 \times 3$ CKM
matrix \cite{pdg}. However, the assumption of the unitarity of the $3
\times 3$ matrix is clearly invalid in the four-generation scenario,
and relaxing it allows a much larger range of values for the elements
$|V_{tq}|$. For example, a large deviation of $|V_{tb}|$ from unity is
claimed to be possible in the SM4
\cite{Alwall:2006bx,Herrera:2008yf,Bobrowski:2009ng,Chanowitz:2009mz,Eberhardt:2010bm}.

We parametrize the CKM4 with 9 parameters, and perform a combined fit
to these parameters using flavor-physics data.  In addition to the
direct measurements of the CKM4 matrix elements, the fit includes
observables that have small hadronic uncertainties:
(i) $R_{bb}$ and $A_b$ from $Z \to b \bar{b}$,
(ii) $\epsilon_K$ from $K_L \to \pi \pi$,
(iii) the branching ratio of $K^+ \to \pi^+ \nu \bar{\nu}$,
(iv) the mass differences in the $B_d$ and $B_s$ systems,
(v) the time-dependent CP asymmetry in $B_d \to J/\psi K_S$,
(vi) the measurement of the angle $\gamma$ of the unitarity triangle
from tree-level decays,
(viii) the branching ratios of $B \to X_s \gamma$ and 
$B \to X_c e \bar{\nu}$, and
(ix) the branching ratio of $B \to X_s \mu^+ \mu^-$ in the
high-$q^2$ and low-$q^2$ regions.
We do not include the oblique parameters in the fit, but simply take
the values of the fourth-generation quark masses to be consistent with
the EW precision data.

There have been several analyses of CKM4 in the past (e.g.\ see
Refs.~\cite{Soni:2010xh,Buras:2010pi,Hou:2010mm,Bobrowski:2009ng,Eberhardt:2010bm,Nandi:2010zx}).
However, they all have a number of deficiencies compared to the
present work.  They do not perform a fit.  Instead, at best, they
present scatter plots showing the allowed ranges of the CKM4 matrix
elements (or correlations between various observables). Of course,
these plots cannot quantify what the errors on the elements are, nor
the confidence level of the ranges. This information can be obtained
only by performing a true fit.  Also, some of them do not include all
clean observables which can be affected by the fourth generation.

The paper is organized as follows.  In Sec.~\ref{ckm4sm4}, we define
the Dighe-Kim parametrization of the CKM4 matrix. In
Sec.~\ref{cons:ckm4}, we present the observables which constrain the
elements of CKM4, along with their experimental values. The results of
the fit are presented in Sec.~\ref{results}. We conclude in
Sec.~\ref{concl} with a discussion of the results.

\section{CKM4 matrix: Dighe-Kim parametrization}
\label{ckm4sm4}

The CKM matrix in the SM is a $3\times3$ unitary matrix:
\beq
V_{\rm CKM3}=\left(\begin{array}{ccc}
V_{ud}& V_{us}&V_{ub}\\ V_{cd}&V_{cs}& V_{cb}\\
V_{td}&V_{ts}&V_{tb}
\end{array}\right)~.
\eeq
In the SM4, the CKM4 matrix is $4\times4$, and can be written as
\beq\label{standard}
V_{\rm CKM4}=\left(\begin{array}{cccc}
\V_{ud}& \V_{us}&\V_{ub}&\V_{ub'}\\ \V_{cd}&\V_{cs}& \V_{cb}&\V_{cb'}\\
\V_{td}&\V_{ts}&\V_{tb}&\V_{tb'}\\\V_{t'd}&\V_{t's}&\V_{t'b}&\V_{t'b'}
\end{array}\right)~.
\eeq
The above matrix can be described, with appropriate choices for the
quark phases, in terms of 6 real quantities and 3 phases.

In this paper, we use the Dighe-Kim (DK) parametrization of the CKM4
matrix \cite{Kim:2007zzg, Alok:2008dj}.  This allows us to treat the
effects of the fourth generation perturbatively and explore the
complete parameter space available. The DK parametrization defines
\beq
\begin{tabular}{lll}
$\V_{us}  \equiv  \lambda$ , &
$\V_{cb}  \equiv A \lambda^2$ , &
$\V_{ub}  \equiv  A \lambda^3 C e^{-i\dub}$ , \\
$\V_{ub'} \equiv  p \lambda^3 e^{-i\dubp}$ , &
$\V_{cb'}  \equiv q \lambda^2 e^{-i\dcbp}$ , &
$\V_{tb'}  \equiv r \lambda $ ~,
\end{tabular}
\label{ckm4tab}
\eeq
where $\lambda$ is the sine of the Cabibbo angle, so that the CKM4 matrix takes the form
\beq
V_{\rm CKM4}=\left(\begin{array}{cccc}
\phantom{sp}\# \phantom{sp} &\phantom{sp} \lambda \phantom{sp} &
A \lambda^3 C e^{-i\dub}&p \lambda^3 e^{-i\dubp}\\ 
 \# & \# & A \lambda^2&q \lambda^2 e^{-i\dcbp}\\
\# & \# & \# &r \lambda\\
\# & \# & \# & \# \\
\end{array}\right) \; .
\label{ckm4dk}
\eeq
The elements denoted by ``$\#$'' can be determined uniquely from the
unitarity condition $V_{\rm CKM4}^\dagger V_{\rm CKM4} = I$.  They can
be calculated in the form of an expansion in powers of $\lambda$ such
that each element is accurate up to a multiplicative factor of $[1 +
  {\cal O}(\lambda^3)]$.

The matrix elements $\V_{ud}$, $\V_{cd}$ and $\V_{cs}$ retain their
SM values:
\beq
\V_{ud} = 1 - \frac{\lambda^2}{2} + {\cal O}(\lambda^4) ~,~~
\V_{cd}  =  -\lambda + {\cal O}(\lambda^5) ~,~~
\V_{cs}  =  1 - \frac{\lambda^2}{2}+ {\cal O}(\lambda^4)\;,
\label{vud} 
\eeq
whereas the values of the matrix elements $V_{td}$, $V_{ts}$ and
$V_{tb}$ are modified due to the presence of the additional quark
generation:
\bea
\V_{td} & = & A \lambda^3 \left( 1 - C e^{i\dub} \right)
+ r \lambda^4 \left( q e^{i\dcbp} - p e^{i\dubp} \right) \nn \\
& & ~~~~~~~
+~\frac{A}{2} \lambda^5 \left( -r^2 + (C + C r^2) e^{i\dub} \right)
+ {\cal O}(\lambda^6) \; , \nn\\
\V_{ts} & = & -A \lambda^2 - q r \lambda^3 e^{i\dcbp}
+ \frac{A}{2} \lambda^4 \left( 1 + r^2 - 2 C e^{i\dub} \right)
+ {\cal O}(\lambda^5) \; , \nn\\
\V_{tb} & = & 1 - \frac{r^2 \lambda^2}{2}+ {\cal O}(\lambda^4) \;.
\label{vtb}
\eea
In the limit $p=q=r=0$, only the elements present in the $3\times 3$
CKM matrix retain nontrivial values, and the above expansion
corresponds to the Wolfenstein parametrization \cite{Wolfenstein} with
$C= \sqrt{\rho^2 + \eta^2}$ and $\dub = \tan^{-1}(\eta/\rho)$.

The remaining new CKM4 matrix elements are:
\bea 
\V_{t'd} & = & \lambda^3 \left( q e^{i\dcbp} - p e^{i\dubp}
\right) + A r \lambda^4 \left( 1 + C e^{i\dub} \right)
\nn \\
& & ~~~~~~~ +~\frac{\lambda^5}{2} \left(p e^{i\dubp} - q r^2
e^{i\dcbp} + pr^2 e^{i\dubp} \right)+ {\cal O}(\lambda^6) \;
, \nn\\
\V_{t's} & = & q \lambda^2 e^{i\dcbp} + A r \lambda^3
\nn \nn\\
& & ~~~~~~~ +~\lambda^4 \left( -p e^{i\dubp} + \frac{q}{2}
e^{i\dcbp} + \frac{q r^2}{2} e^{i\dcbp} \right)+ {\cal
O}(\lambda^5) \; , \nn\\
\V_{t'b} & = & - r \lambda+ {\cal O}(\lambda^4) \; , \nn\\
\V_{t'b'} & = & 1 - \frac{r^2 \lambda^2}{2} + {\cal
O}(\lambda^4) \; .
\eea

\section{Constraints on the CKM4 matrix elements}
\label{cons:ckm4}

In order to obtain constraints on the CKM4 matrix elements, we perform
a $\chi^2$ fit for all nine CKM4 parameters using the CERN
minimization code MINUIT \cite{minuit}.  The fit is carried out for
$m_{t'}=400$ GeV and $600$ GeV.  The $b'$ mass is fixed by the
relation $m_{t'} - m_{b'} = 55\,\rm GeV$ [see
  Eq.~(\ref{EWconstraints})].  We include both experimental errors and
theoretical uncertainties in the fit. In the following subsections, we
discuss the various observables used as constraints, and give their
experimental values.

\subsection{Direct Measurements of the CKM Elements}

The values of CKM elements obtained from the measurement of the
tree-level weak decays are independent of the number of generations.
Hence they apply to the $3 \times 3$ and $4 \times 4$ matrices.
The elements $|\V_{ud}|$, $|\V_{us}|$, $|\V_{ub}|$, $|\V_{cd}|$, $|\V_{cs}|$ 
and $|\V_{cb}|$ have all been directly measured. 
We use the following measurements \cite{pdg}
to constrain the CKM4 parameters:
\bea
|\V_{ud}| = (0.97418 \pm 0.00027)\;, & & 
|\V_{cd}| = (0.23 \pm 0.011)\;,  \nn\\
|\V_{us}| = (0.2255 \pm 0.0019)\;, & &
|\V_{cs}| = (1.04 \pm 0.06)\;,  \nn\\
|\V_{ub}| = (3.93 \pm 0.36)\, \times 10^{-3}\;, & &
|\V_{cb}| = (41.2 \pm 1.1)\, \times 10^{-3}\;.
\eea

\subsection{Unitarity of the CKM4 Matrix}

Constraints on the CKM4 matrix elements can be obtained by using the
unitarity of the CKM4 matrix. Through a variety of independent
measurements, the SM $3 \times 3$ submatrix has been found to be
approximately unitary. We therefore expect all the CKM4 matrix
elements which involve both the fourth-generation and light quarks to
be relatively small.

Using the measurements of $|V_{ud}|$, $|V_{us}|$ and $|V_{ub}|$, the
first row of the CKM4 matrix gives
\beq
|\V_{ub'}|^2 = 1- \big(|\V_{ud}|^2+|\V_{us}|^2+|\V_{ub}|^2\big)
= 0.0001 \pm 0.0011\;.
\eeq
Using the measurements of $|\V_{cd}|$, $|\V_{cs}|$ and $|\V_{cb}|$, the
second row gives
\beq
|\V_{cb'}|^2 = 1- \big(|\V_{cd}|^2+|\V_{cs}|^2+|\V_{cb}|^2\big)
= -0.136 \pm 0.125\;.
\eeq
Similarly, from the first column of CKM4, we have
\beq
|\V_{td}|^2 + |\V_{t'd}|^2 = 1- \big(|\V_{ud}|^2+|\V_{cd}|^2\big)
= -0.002 \pm 0.005\;.
\eeq
Finally, the second column of CKM4 implies
\beq
|\V_{ts}|^2 + |\V_{t's}|^2 = 1- \big(|\V_{us}|^2+|\V_{cs}|^2\big)
= -0.134 \pm 0.125\;.
\eeq

\subsection{\boldmath Vertex Corrections to $Z\to b\bar{b}$} 

Including the QCD and QED corrections, the decay rate for $Z\to
b\bar{b}$ is given by \cite{Bernabeu}
\bea
\Gamma(Z\to q\bar{q}) &=& {\frac{\alpha\; m_Z} {16\sin^2\theta_W \cos^2\theta_W}} 
\left(|a_q|^2 + |v_q|^2\right) \big(1 + \delta_q^{(0)}\big)
\nonumber \\&{}& 
\times\big(1 + \delta^q_{QED}\big)\big(1 + \delta^q_{QCD}\big)
\big(1 + \delta^q_{\mu}\big)\big(1 + \delta^q_{tQCD}\big)\big(1 + \delta_b \big) \;.
\eea
Here $a_q = 2 I^q_3$ and $v_q = \Big(2 I^q_3 - 4 |Q_q|
\sin^2\theta_W\Big)$ are the axial and vector coupling constants,
respectively.

The $\delta$ terms are corrections due to various higher-order loops:
\begin{itemize}

\item $\delta_q^{(0)}$ contains small electroweak corrections not
  absorbed in $\sin^2\theta_W$. Their effect is at most at the 0.5\%
  level.

\item $\delta^{q}_{QED}$ represents small final-state QED corrections
  that depend on the charge of final fermion.  It is very small: 0.2\%
  for the charged leptons, 0.8\% for the $u$-type quarks and 0.02\% for
  the $d$-type quarks \cite{Bernabeu}.

\item $\delta_{QCD}$ includes the QCD corrections common to all
  quarks; it is given by \cite{Bernabeu}
\beq
\delta_{QCD} = \frac{\alpha_s}{\pi}  + 1.41 \Big(\frac{\alpha_s}{\pi} \Big)^2\;,
\eeq
where $\alpha_s$ is the QCD coupling constant taken at the $m_Z$
scale: $\alpha_s= \alpha_s(m^2_Z) = 0.12$.

\item $\delta^q_{\mu}$ contains the kinematical effects of the
  external fermion masses, including some mass-dependent QCD radiative
  corrections. It is only important for the $b$ quark (0.5\%), and to
  a lesser extent for the $\tau$ lepton (0.2\%) and the $c$ quark
  (0.05\%) \cite {Bernabeu,Chetyrkin:1990kr}.

\item The correction $\delta^q_{tQCD}$ consists of QCD contributions
  to the axial part of the decay and originates from doublets with
  large mass splitting \cite{Bernabeu,Kniehl:1989bb}. In the presence
  of the fourth generation, it is given by \cite{Yanir:2002cq}
\beq
\delta^q_{tQCD} = -\frac{a_q}{v^2_q + a^2_q} \Big(\frac{\alpha_s}{\pi} \Big)^2
\Bigg[a_t\,f(\mu_t)+a_{t'}\,f(\mu_{t'})+a_{b'}\,f(\mu_{b'})\Bigg]\;,
\label{tqcd}
\eeq
where
\beq
f(\mu_f)\approx \log\big(\frac{4}{\mu^2_f}\big)-3.083+\frac{0.346}{\mu^2_f}+\frac{0.211}{\mu^4_f}\;,
\eeq
with $\mu^2_f=4m^2_f/m^2_Z$.

\item $\delta_b$ is non-zero only for $q=b$ and is due to the
  $Zb\bar{b}$ vertex loop corrections.  In the presence of the fourth
  generation, it is given by \cite{Yanir:2002cq}
\beq
\delta_b \approx  10^{-2}\left[\left(-\frac{m^2_t}{2 m^2_Z} + 0.2\right)
|\V_{tb}|^2
 + \left(- \frac{m^2_{t'}}{2 m^2_Z} + 0.2\right)|\V_{t'b}|^2\right]\;.
\label{deltab}
\eeq

\end{itemize}

In order to isolate the large mass dependences appearing in the
$Zb\bar{b}$ vertex $\delta_b$, one takes the following ratio
\cite{Bernabeu}:
\beq
R_{bb}  = \big( 1 + \frac{2}{R_s}
+ \frac{1}{R_c} + \frac{1}{R_u} \big)^{-1}\;,
\eeq
where
\bea
R_{s} & \equiv & \frac{\Gamma(Z\to b\bar{b})}{\Gamma(Z\to s\bar{s})} \approx 0.9949\big(1 + \delta_b \big)\;,
\label{rs}
\nn\\
R_c & \equiv & \frac{\Gamma(Z\to b\bar{b})}{\Gamma(Z\to c\bar{c})} \approx 0.9960\frac{\big(1+v^2_b\big)}{\big(1+v^2_c\big)}\frac{\big(1+\delta^b_{tQCD}\big)}{\big(1+\delta^c_{tQCD}\big)}\big(1 + \delta_b \big)\;.
\label{rc}
\nn\\
R_u & \equiv & \frac{\Gamma(Z\to b\bar{b})}{\Gamma(Z\to u\bar{u})} \approx 
0.9955\frac{\big(1+v^2_b\big)}{\big(1+v^2_c\big)}\frac{\big(1+\delta^b_{tQCD}\big)}{\big(1+\delta^c_{tQCD}\big)}\big(1 + \delta_b \big)\;.
\label{ru}
\eea
Using Eqs.~(\ref{tqcd})-(\ref{rc}), we get (for $m_{t'}= 400$-$600$ GeV)
\beq
R_{bb}=\Bigg[1+\frac{3.584}{\big(1 + \delta_b \big)}\Bigg]^{-1}\;.
\label{rbbf}
\eeq
The data give \cite{lepewwg}
\beq
R_{bb} = 0.216 \pm 0.001 \; ,
\eeq
which, through Eq.~(\ref{deltab}), determines a linear combination of
$|\V_{tb}|^2$ and $|\V_{t'b}|^2$.
This constrains the combination $r \lambda$ of the CKM4 parameters.

We also consider constraints from the forward-backward (FB) asymmetry
in $Z\to b\bar{b}$.  The $Z\to b\bar{b}$ interaction Lagrangian is
\beq
{\cal L} = \frac{g}{\cos \theta_W} \bar{b}\gamma^{\mu}\Big(g_{bL}P_L + g_{bR}P_R\Big)b\,Z_{\mu}\,,
\eeq
where $P_{L(R)}$ are the chirality projection operators, and 
\bea
g_{bL} &=& -\frac{1}{2} + \frac{1}{3} \sin^2 \theta_W + \delta g_{bL}^t + \delta g_{bL}^{t'}\,,
\\
g_{bR} &=&  \frac{1}{3} \sin^2 \theta_W + \delta g_{bR}^t + \delta g_{bR}^{t'}\,.
\eea
Here the $\delta$'s represent the radiative corrections due to the $t$
and $t'$ quarks.  The FB asymmetry in $Z\to b\bar{b}$ 
allows us to determine the asymmetry parameter\footnote{
The measured FB asymmetry $A_{FB}^{0,b}$ is related to the asymmetry
parameter $A_b$ via $A_{FB}^{0,b} \approx (3/4) A_e A_b$, where $A_e$ 
is the corresponding asymmetry parameter for the electron \cite{vysotsky}.
We only consider the parameter $A_b$ since $A_e$, and hence $A_{FB}^{0,b}$
itself, would involve contribution from the fourth generation lepton sector,
while we would like to restrict ourselves to the quark sector in this paper.}
\beq
A_b = \frac{g_{bL}^2 - g_{bR}^2}{g_{bL}^2 + g_{bR}^2}\,.
\label{zafb}
\eeq

Within both the SM and SM4, only the $g_{bL}$ terms receive corrections
proportional to $m_{t,t'}^2$ at the loop level. We have
\cite{Chanowitz:2009mz,Akhundov:1985fc,Beenakker:1988pv,Bernabeu:1987me,Lynn:1990hd}
\bea
\delta g_{bL}^t&=&\frac{\alpha}{16 \pi \sin^2 \theta_W  \cos^2 \theta_W }\frac{m^2_t}{ m^2_Z}|\V_{tb}|^2\,,
\\
\delta g_{bL}^{\rm t'}&=&\frac{\alpha}{16 \pi \sin^2 \theta_W  \cos^2 \theta_W }\frac{m^2_{t'}}{ m^2_Z}|\V_{t'b}|^2\,,
\\
g_{bR}^t &=&0\,,
\\
g_{bR}^{\rm t'}&=&0\,.
\eea
The data give \cite{lepewwg}
\beq
A_b = 0.923\pm0.020 \; ,
\eeq
which, through Eq.~(\ref{zafb}), constrains the combination $r
\lambda$ of the CKM4 parameters.

\subsection{\boldmath The $K$ system}

Here we present observables in various $K$ decays with the addition of
a fourth generation.

\subsubsection{Indirect CP violation in $K_L \to \pi\pi$}

Indirect CP violation in $K_L\to \pi \pi$ is described by the
parameter $\epsilon_K$, given by \cite{Buras:2010pi,Buras:2008nn}
\beq
\epsilon_K = \frac{k_\epsilon e^{i \phi_{\epsilon}}}{\sqrt{2}(\Delta M_K)_{\rm exp}} {\rm Im} (M^K_{12})\;.
\label{epsilonk}
\eeq
$(\Delta M_K)_{\rm exp}$ is the $K_L$-$K_S$ mass difference. The
parameters $\phi_{\epsilon}=(43.51 \pm 0.05)^{\circ}$ and
${\kappa_{\epsilon}}= 0.92\pm0.02$ \cite{Buras:2008nn} include an
additional effect from ${\rm Im}({A_0})$, where $A_0\equiv A\big(K\to
(\pi\pi)_{I=0}\big)$. $M^K_{12}$ is the off-diagonal element in the
dispersive part of the amplitude for $K^0$-$\bar{K^0}$ mixing:
\beq
\big(M^{K}_{12}\big)^{*} = \frac{{\bra\bar{K^0}|{\cal{H}}^{\Delta S = 2}_{eff}|K^0\ket}}{2 m_K} ~.
\eeq

The calculation of $M^K_{12}$ in the SM4 gives \cite{Buras:1997fb}
\bea
M^{K}_{12} & = & \left(\frac{G^2_F M^2_W}{12\pi^2}\right) m_K \hat{B}_K f^2_K 
\Big[\eta_c (\V^*_{cd}\V_{cs})^2 S(x_c) 
+ 2 \eta_{ct} (\V^*_{cd}\V_{cs}) (\V^*_{td}\V_{ts}) S(x_c, x_t)
\nn \\ &&
+~\eta_t (\V^*_{td}\V_{ts})^2 S(x_t)
+ 2 \eta_{ct'} (\V^*_{cd}\V_{cs}) (\V^*_{t'd}\V_{t's}) S(x_c, x_{t'})
\nn \\ &&
+~2 \eta_{tt'} (\V^*_{td}\V_{ts}) (\V^*_{t'd}\V_{t's}) S(x_t, x_{t'})
+ \eta_{t'} (\V^*_{t'd}\V_{t's})^2 S(x_{t'})\Big]\;.
\label{m12k}
\eea
For the decay constant and bag parameter we take $f_K = (155.8 \pm
1.7)\,{\rm MeV}$ \cite{Laiho:2009eu}, $\hat{B}_K = 0.725 \pm 0.026$
\cite{Laiho:2009eu}.  The Inami-Lim functions $S(x)$ and $S(x,y)$ are
\cite{Inami:1980fz}
\bea
S(x) &=& \frac{4 x - 11 x^2 + x^3}{4 (1-x)^2}
-\frac{3}{2} \frac{x^3 \mbox{\rm ln} x}{(1-x)^3}\;,\nn\\
S(x,y) &=& x y \Bigg\{\frac{\ln y}{y-x} \Bigg[
\frac{1}{4} + \frac{3}{2} \frac{1}{1-y} -
\frac{3}{4} \frac{1}{(1-y)^2} \Bigg]  
 \nn \\
&& -~\frac{\ln x}{y-x} \Bigg[\frac{1}{4} + \frac{3}{2} \frac{1}{1-x} -
\frac{3}{4} \frac{1}{(1-x)^2} \Bigg]  
 -\frac{3}{4} \frac{1}{(1-x) (1-y)} \Bigg\}\;,
 \label{inami-lim}
\eea
where $x=m^2_q/M^2_W$ for all quarks $q$.

The predictions for the short-distance QCD factors are:
$\eta_c=(1.51\pm0.24)$ \cite{Herrlich:1993yv}, $\eta_{ct}=0.47\pm
0.04$ \cite{Herrlich:1995hh, Herrlich:1996vf}, $\eta_t=0.58$
\cite{Buras:1990fn}.  The values for $\eta_{ct}$ and $\eta_c$ have a
sizeable uncertainty as they are sensitive to the light scale $\sim
m_c$ where $\alpha_s$ is large. The QCD correction factor $\eta_{t'}$
is given by \cite{Hattori:1999ap}
\bea
\eta_{t'} = \Big(\alpha_s(m_t)\Big)^{6/23} 
\left( \frac{\alpha_s(m_{b^\prime})}{\alpha_s(m_t)} \right)^{6/21} 
\left( \frac{\alpha_s(m_{t^\prime})}{\alpha_s(m_{b^\prime})} \right)^{6/19}\;.
\eea
$\alpha_s(\mu)$ is the running coupling constant at the scale
$\mu$ at NLO \cite{Buchalla:1995vs}.  Here we assume
$\eta_{tt'}=\eta_{t'}$ and $\eta_{ct'}=\eta_{ct}$.

Using Eqs.~(\ref{epsilonk}) and (\ref{m12k}), we get
\bea
\epsilon_K  & = & \frac{G^2_F M^2_W f^2_K m_K \hat{B}_K k_\epsilon e^{i \phi_{\epsilon}} }{12\sqrt{2}\pi^2(\Delta M_K)_{\rm exp}}  
{\rm Im}\Big[\eta_c (\V^*_{cd}\V_{cs})^2 S(x_c) 
+ 2 \eta_{ct} (\V^*_{cd}\V_{cs}) (\V^*_{td}\V_{ts}) S(x_c, x_t)
\nn \\ &&
+~\eta_t (\V^*_{td}\V_{ts})^2 S(x_t)
+ 2 \eta_{ct'} (\V^*_{cd}\V_{cs}) (\V^*_{t'd}\V_{t's}) S(x_c, x_{t'})
\nn \\ &&
+~2 \eta_{tt'} (\V^*_{td}\V_{ts}) (\V^*_{t'd}\V_{t's}) S(x_t, x_{t'})
+ \eta_{t'} (\V^*_{t'd}\V_{t's})^2 S(x_{t'})\Big]\;.
\label{cpvkpipi}
\eea
The measured value is $|\epsilon_K| = (2.32 \pm 0.007) \times 10^{-3}$
\cite{pdg}. This which mainly puts constraints on the combinations
$\V_{td}^* \V_{ts}$ and $\V_{t'd}^* \V_{t's}$, which, to leading order
in $\lambda$, depend on $A^2 [1 - C e^{i\delta_{ub}}]$ and $q^2 [1 -
  (p/q) e^{i (\delta_{cb'} -\delta_{ub'})}]$, respectively.

\subsubsection{ $K^+\to \pi^+\nu \bar{\nu}$}

The flavor-changing neutral-current (FCNC) quark-level transition
${\bar s} \to {\bar d} \nu \bar{\nu}$ is responsible for the decay
$K^+\to \pi^+\nu \bar{\nu}$.  Unlike other $K$ decays, $K^+\to
\pi^+\nu \bar{\nu}$ is dominated by the short-distance (SD)
interactions. The long-distance (LD) contribution to $Br(K^+\to
\pi^+\nu \bar{\nu})$ is about three orders of magnitude smaller than
that of the SD \cite{Rein:1989tr, Hagelin:1989wt}.  As the decay
$K^+\to \pi^+\nu \bar{\nu}$ occurs via loops containing virtual heavy
particles, it is sensitive to the fourth-generation quark $t'$.

The effective Hamiltonian for the decay $K^+\to \pi^+\nu \bar{\nu}$ in
the SM4 can be written as
\bea
{\cal{H}}_{eff} &=& {G_F \over \sqrt{2}}{\alpha\over {2\pi s^2_W}} 
\sum_{l=e, \mu, \tau}\Big[ \V_{cs}^{*}\V_{cd} X^l_{NL} 
+ \V_{ts}^{*}\V_{td} X(x_t) 
\nn \\ &&
~~~~~~ + \V_{t's}^{*}\V_{t'd} X(x_{t'}) \Big](\bar{s}d)_{V-A}(\bar{\nu}_l\nu_l)_{V-A}\;.
\eea
The function $X(x)$ ($x \equiv m^2_{t,t'}/M^2_W$), relevant for the
$t$ and $t'$ pieces, is given by
\beq
X(x) = \eta_X X_0(x)\;,
\eeq
where 
\beq
X_0(x) = {x\over 8}\left[- {2 + x \over 1-x} + {3x-6 \over (1-x)^2} \ln{x}\right]\;.
\eeq
Above, $\eta_X$ is the NLO QCD correction; its value is estimated to
be $0.994$ \cite{Buras:1997fb}. The function corresponding to $X(x)$
in the charm sector is $X^l_{NL}$:
\beq
X^l_{NL}=C_{NL}-4B^{(1/2)}_{NL}\;,
\eeq
where $C_{NL}$ and $B^{(1/2)}_{NL}$ correspond to the
electroweak-penguin and box contributions, respectively. The explicit
forms of $C_{NL}$ and $B^{(1/2)}_{NL}$ are given in
Refs.~\cite{Buchalla:1995vs, Buchalla:1993wq}.

The branching ratio of $K^+\to \pi^+\nu\bar{\nu}$ in the SM4 is given
by
\bea
Br(K^+\to \pi^+\nu\bar{\nu}) &=& \kappa_+ \Big[\left({{\rm Im}(\V_{ts}^{*}\V_{td}) \over \lambda^5}X(x_t) 
+ {{\rm Im}(\V_{t's}^{*}\V_{t'd}) \over \lambda^5} X(x_{t'}) \right)^2 
\nn \\ 
&& \hskip0.7truecm
 +~\Bigg( \frac{{\rm Re}(\V_{cs}^{*}\V_{cd})}{\lambda}P_0(X) 
+ \frac{{\rm Re}(\V_{ts}^{*}\V_{td})}{\lambda^5}X(x_t) 
\nn \\ 
&& \hskip0.7truecm
+~\frac{{\rm Re}(\V_{t's}^{*}\V_{t'd})}{\lambda^5} X(x_{t'})\Bigg)^2 \Big],
\label{brkpinunu}
\eea 
where 
\beq
\kappa_+ = r_{K+} {{3\alpha^2 \,Br(K^+\to \pi^0 e^+ \nu )}\over {2\pi^2 s^4_W}}\lambda^8 ~~,~~~~
P_0(X) = {1\over\lambda^4}\left[{2\over 3} X^e_{NL} + {1\over 3} X^{\tau}_{NL}\right]\;.
\eeq
We see that $Br(K^+\to \pi^+\nu \bar{\nu})$ is related to the
experimentally well-known quantity $Br(K^+\to \pi^0 e^+ \nu)$.
$r_{K+}$ summarizes the isospin-breaking corrections in relating
$K^+\to \pi^+\nu\bar{\nu}$ to $K^+\to \pi^0 e^+ \nu $; its value is
$r_{K+} = 0.901$.  $\kappa_+ $ is estimated to be $(5.36\pm0.026)
\times 10^{-11}$ \cite{Mescia:2007kn}.

The measured value is $Br(K^+\to \pi^+\nu{\bar\nu}) =
(1.7\pm1.1)\times 10^{-10}$ \cite{pdg}. This which mainly puts
constraints on the combinations $\V_{td}^* \V_{ts}$ and $\V_{t'd}^*
\V_{t's}$, which, to leading order in $\lambda$, depend on $A^2 [1 - C
  e^{i\delta_{ub}}]$ and $q^2 [1 - (p/q) e^{i (\delta_{cb'}
    -\delta_{ub'})}]$, respectively.

\subsection{\boldmath The $B$ system}

Here we present various observables in the $B$ system with the
addition of a fourth generation.

\subsubsection {  $B_{d,s}^0$-$\bar{B}_{d,s}^0$ mixing}

Meson-antimeson mixing occurs in the SM through the box diagram, and
is thus sensitive to new heavy particles appearing in the loop.
Within the three-generation SM, the dominant contribution to
$B_{q}^0$-$\bar B_{q}^0$ mixing ($q=d,s$) comes from the virtual top
quark.  The charm and the mixed top-charm contributions are negligibly
small, and hence the analysis is simplified considerably.  In the SM4,
there is an additional contribution due to the virtual $t'$ in the box
diagram.

The mass difference $\Delta M_q$ is given by $\Delta M \simeq
2|M^q_{12}|$, where $M^q_{12}$ is the virtual part of the box diagrams
responsible for the mixing. For $B_{q}^0$-$\bar B_{q}^0$ mixing,
$M^q_{12}$ in the SM4 is given by
\bea
M_{12}^{q} &=& \frac{G^2_F M_W^2}{12\pi^2} m_{B_q} \hat{B}_{bq} f_{B_q}^2 \Big[
\eta_t (\V^{*}_{tq} \V_{tb})^2 S(x_t)   +
\eta_{t'} (\V^{*}_{t' q} \V_{t' b})^2 S(x_{t'}) 
\nn \\&& 
~~~~~~~ +~2\eta_{tt'} (\V^{*}_{tq} \V_{tb})\, ( \V^{*}_{t' q} \V_{t' b}) 
S(x_t,x_{t'}) \Big]\;,
\label{M12q}
\eea
where $x_t=m_t^2/m_W^2$, $x_{t'}=m_{t'}^2/M_W^2$. The Inami-Lim
functions $S(x)$ and $S(x,y)$ are given in Eq.~(\ref{inami-lim}).
Here we assume $\eta_{t'}=\eta_{tt'}$ for simplicity. The numerical
values of the structure functions $S(x_{t'})$, $S(x_t,x_{t'})$ and the
QCD correction factor $\eta_{t'}$ for various $t'$ mass are given in
Ref.~\cite{Soni:2010xh}. In order to reduce the sizeable
non-perturbative uncertainties due to the decay constant $f_{B_q}$ and
the bag parameter $\hat{B}_{bq}$, we consider the ratio $\Delta
M_s/\Delta M_d$:
\beq
\frac{\Delta M_s}{\Delta M_d}  =  \frac{m_{B_s}}{m_{B_d}}\, \xi^2 \times M_R\;,\\
\eeq
where
\beq
M_R=\frac{\left|\eta_t (\V^{*}_{ts} \V_{tb})^2 S(x_t)   +
\eta_{t'} (\V^{*}_{t' s} \V_{t' b})^2 S(x_{t'}) 
+~2\eta_{tt'} (\V^{*}_{ts} \V_{tb})\, ( \V^{*}_{t' s} \V_{t' b}) S(x_t,x_{t'})
\right|}{\left|\eta_t (\V^{*}_{td} \V_{tb})^2 S(x_t)   +
\eta_{t'} (\V^{*}_{t' d} \V_{t' b})^2 S(x_{t'}) 
+~2\eta_{tt'} (\V^{*}_{td} \V_{tb})\, ( \V^{*}_{t' d} \V_{t' b}) S(x_t,x_{t'})\right|}\;,
\label{eq:mr}
\eeq
with $\xi \equiv {f_{B_s}\sqrt{\hat{B}_{bs}}} /
{f_{B_d}\sqrt{\hat{B}_{bd}}} = 1.243 \pm 0.028$
\cite{Laiho:2009eu}. There is less uncertainty in $\xi$ ($\sim
2$-$3\%$) than in $f_{B_q}\sqrt{\hat{B}_{bq}}$ ($\sim 7$-$8\%$).

The measured values are \cite{pdg}
\beq
\Delta{M_s} = (17.77 \pm 0.12)\, {\rm ps}^{-1} ~~,~~~~
\Delta{M_d} = (0.507 \pm 0.005) \, {\rm ps}^{-1} ~,
\eeq
whose ratio is sensitive to $\V_{tb}^* \V_{ts}$, $\V_{tb}^* \V_{td}$,
$\V_{t'b}^* \V_{t's}$, and $\V_{t'b}^* \V_{t'd}$.  These correspond to
the combinations of the CKM4 parameters $A^2$, $A^2 (1 - C
e^{i\delta_{ub}})$, $q r e^{i \delta_{cb'}}$, and $r (q e^{i
  \delta_{cb'}} - p e^{i \delta_{ub'}})$, respectively, to leading
order in $\lambda$.

\subsubsection {CP violation}

CP violation in the quark sector is due to phases in the quark mixing
matrix. In the three-generation SM, the phase information in the CKM
matrix is elegantly encapsulated in the unitarity triangle \cite{pdg},
whose interior angles are $\alpha$, $\beta$ and $\gamma$. In order to
test the SM, these angles must be measured in as many ways as possible
to test for consistency. Unknown strong QCD phases contaminate many of
these methods; ways of removing these strong phases must be devised in
order to cleanly measure the weak phases. In the SM4, many of the
ways of eliminating the strong phases fail, since typically
there are multiple amplitudes with different strong and weak phases. 
Here we consider only those constraints from CP observables 
which are free from uncertainties due to the strong phases.

\begin{itemize}

\item $S_{J/\psi K_S}$: The coefficient of $\sin(\Delta M_d t)$ in the
  time-dependent indirect CP asymmetry in $B_{d}^0 \to J/\psi K_S$ is
  given by
\beq
S_{J/\psi K_S}=\sin 2\phi^{\rm tot}_{B_{d}}\;,
\eeq
where $\phi^{\rm tot}_{B_{d}}$ is defined as
\beq
M_{12}^{d}=|M_{12}^{d}| e^{i\,2\phi^{\rm tot}_{B_{d}}}\;.
\eeq
Thus we have
\beq
S_{J/\psi K_S}=\frac{{\rm Im}(M_{12}^{d})}{|M_{12}^{d}|}\;.
\label{eq:sin2b}
\eeq
In the SM, this is $\sin 2\beta$, and is free of strong phases. It is
thus a good observable to constrain the SM4 using Eq.~(\ref{M12q}).
The measured value is \cite{pdg}
\beq
S_{J/\psi K_S}=0.672\pm0.024 ~,
\eeq
which is sensitive to $\V_{tb}^* \V_{td}$ and $\V_{t'b}^* \V_{t'd}$,
i.e. to the parameter combinations
$A^2 (1 - C e^{i\delta_{ub}})$ and 
$r (q e^{i \delta_{cb'}} - p e^{i \delta_{ub'}})$, respectively.

\item Recently, the CDF and D0 collaborations measured indirect CP
  violation in $\bs \to J/\psi \phi$ and found a 2.2$\sigma$ deviation
  from the prediction of the SM \cite{CDFD0}. At first sight, this
  seems to indicate a nonzero phase of $\bs$-$\bsbar$ mixing, and many
  papers have been written exploring the contribution of particular
  new-physics models to this mixing (including the fourth generation
  \cite{Soni:2008bc}).  However, there could be a significant
  contribution to this signal from new physics in the decay
  \cite{BsNPdecay}.  If this is the case, strong phases will play a
  role. For this reason, the constraints from this measurement are not
  included in the fit.

\item In the SM, $\gamma \equiv {\rm Arg}( - V_{ub}^*
  V_{ud})/(V_{cb}^* V_{cd})$. This phase can be probed in tree-level
  decays. By measuring several different decays, it is possible to
  remove the dependence on the strong phase and extract $\gamma$. The
  latest value is \cite{pdg}
\beq
\gamma=(75.0 \pm 22.0)^{\circ} ~.
\eeq

Because this angle is measured in tree-level decays, its value is
unchanged with the addition of a fourth generation. Indeed, from
Eqs.~(\ref{ckm4tab}) and (\ref{vud}), we see that
\beq
{\rm Arg}\left(- \frac{V_{ub}^* V_{ud}}{V_{cb}^* V_{cd}} \right)
\approx {\rm Arg}\left(- \frac{\V_{ub}^* \V_{ud}}{\V_{cb}^* \V_{cd}} \right)
\approx \delta_{ub} \; .
 \eeq
Thus, the phase $\delta_{ub}$ can be constrained through the
measurement of the weak phase $\gamma$, and this observable is
included in the fit.

\end{itemize}

\subsubsection { $B \to X_s \gamma$}

The quark-level transition $\bar{b} \to \bar{s} \gamma$ induces the
inclusive radiative decay $B \to X_s \gamma$. This decay can occur
only at the loop level and hence is suppressed within the SM. It has
been observed with a branching ratio of $(3.55 \pm 0.25)\times
10^{-4}$ \cite{hfag}, in good agreement with the NNLO SM prediction of
$(3.15 \pm 0.23)\times 10^{-4}$ \cite{misiak}. Thus, $B \to X_s
\gamma$ has a great potential to constrain new-physics models.

Within the SM, the effective Hamiltonian for the quark-level
transition $\bar{b} \to \bar{s} \gamma$ can be written as
\beq
{\cal H}_{eff} =  \frac{4 G_F}{\sqrt{2}} V_{ts}  V_{tb}^{*}
\sum_{i=1}^{8} C_i(\mu) \, Q_i(\mu)\;,
\eeq
where the form of the operators $O_i(\mu)$ and the expressions for
calculating the Wilson coefficients $C_i(\mu)$ are given in
Ref.~\cite{Buras:1994dj}.  
The introduction of a fourth generation,
in addition to the modifications $V_{ts} \to \V_{ts}$ and
$V_{tb} \to \V_{tb}$,
also changes the values of the Wilson coefficients $C_{7,8}$ via the
virtual exchange of the $t'$-quark. They can be written as
\beq
C_{7,8}^{\rm tot}(\mu) = C_{7,8}(\mu) + \frac{\V^{*}_{t^{'}b}\V_{t^{'}s}}
{\V^{*}_{tb}\V_{ts}} C_{7,8}^{t'} (\mu)\;.
\label{wtot_78}
\eeq
The values of $C_{7,8}^{t'}$ can be calculated from the expressions
for $C_{7,8}$ by replacing $m_{t}$ by $m_{t'}$.

In order to reduce the large uncertainties arising from $b$-quark
mass, we consider the following ratio 
\beq
R = \frac{Br(B \to X_s \gamma)}{Br(B \to X_c e \bar \nu_e)}\;. \nonumber
\eeq
In leading logarithmic approximation this ratio can be written as 
\cite{Buras:1997fb}
\beq
R = \frac{\left| \V_{tb}^{*} \V_{ts} \right|^2}{\left| \V_{cb} \right|^2} \,\,
\frac{6 \alpha \left| C_7^{\rm tot}(m_b) \right|^2}{\pi f(\hat m_c) 
\kappa(\hat m_c)}\;. 
\label{R}
\eeq
Here the Wilson coefficient $C_7$ is evaluated at the scale $\mu=m_b$.
The phase space factor $f(\hat{m_c})$ in $Br(B \to X_c e {\bar \nu})$ 
is given by \cite{Nir:1989rm}
\beq
f(\hat{m}_c) = 1 - 8\hat{m}^2_c + 8\hat{m}_c^6 - \hat{m}_c^8 - 
24\hat{m}_c^4 \ln \hat{m}_c \;,
\label{fmc}
\eeq
where $\hat{m_c}=m_c/m_b$. $\kappa(\hat{m_c})$ is the $1$-loop QCD
correction factor \cite{Nir:1989rm}:
 \beq
\kappa(\hat{m_c})=1-\frac{2\alpha_s(m_b)}{3\pi}\left[\left(\pi^2-\frac{31}{4}\right)(1-\hat{m_c})^2+\frac{3}{2}\right]\;.
\label{kappa}
\eeq

The values of the branching ratios in $R$ are $Br(B\to X_s \gamma) =
(3.55 \pm 0.25)\times 10^{-4}$ \cite{pdg} and $Br(B \to X_c e
{\bar\nu}) = 0.1061\pm0.0016\pm0.0006$ \cite{Aubert:2004aw}),
the ratio being sensitive to
$\V_{tb}^* \V_{ts}$ and $\V_{t'b}^* \V_{t's}$,
i.e. to $A^2$ and $q r e^{i \delta_{cb'}}$, respectively,
to leading order in $\lambda$.

\subsubsection { $B \to X_s\, l^+ \,l^-$}

The effective Hamiltonian for the quark-level transition
$\bar{b} \to \bar{s}\, l^+\, l^-$ in the SM can be written as
\beq
{\cal H}_{eff} =  \frac{4 G_F}{\sqrt{2}} V_{ts} V_{tb}^{*}
\sum_{i=1}^{10} C_i(\mu) \,  Q_i(\mu)\;,
\eeq
where the form of the operators $Q_i$ and the expressions for
calculating the coefficients $C_i$ are given in
Ref.~\cite{Buras:1994dj}.  The fourth generation,
in addition to the modifications $V_{ts} \to \V_{ts}$ and
$V_{tb} \to \V_{tb}$,
changes the values of the Wilson coefficients $C_{7,8,9,10}$ via the
virtual exchange of the $t^\prime$.  The Wilson coefficients in
the SM4 can then be written as
\beq
C^{\rm tot}_{i}(\mu_b)=C_{i}(\mu_b)
+\frac{\V^{*}_{t^{'}b}\V_{t^{'}s}}
{\V^{*}_{tb}\V_{ts}}C^{t'}_{i}(\mu_b) ~,
\eeq
where $i=7,8,9,10$. The new Wilson coefficients
$C^{t'}_{i}(\mu_b)$ can easily be calculated by substituting
$m_{t^{'}} $ for $m_t$ in the SM expressions involving the
$t$ quark.
 
The calculation of the differential decay rate gives
\beq
\frac{{\rm d}Br(B \to X_s\, l^+ \,l^-)}{{\rm d}z}
 = \frac{\alpha^2 Br(B\rightarrow X_c e {\bar \nu})}
 {4 \pi^2 f(\hat{m_c})\kappa(\hat{m}_c)} 
 \frac{|\V_{tb}^{*}\V_{ts}|^2}{|\V_{cb}|^2} (1-z)^2 D(z)\,,
\label{eq:brincl}
\eeq
where
\beq
D(z) =  (1+2z) \left( |C_9^{\rm tot}|^2 + |C_{10}^{\rm tot}|^2 \right)
      + 4 \left(1+\frac{2}{z}\right) |C_7^{\rm tot}|^2
     +12 {\rm Re}(C_7^{\rm tot} C_{9}^{\rm tot*})\;.
\eeq
Here $z \equiv q^2/m_b^2$ and $\hat{m}_q=m_q/m_b$ for all quarks $q$.
The expressions for the phase-space factor $f(\hat{m_c})$ and the
$1$-loop QCD correction factor $\kappa(\hat{m_c})$ are given
Eqs.~(\ref{fmc}) and (\ref{kappa}), respectively.

The theoretical prediction for the branching ratio of $B \to X_s\, l^+
\,l^-$ in the intermediate $q^2$ region ($7$~GeV$^2 \le q^2 \le
12$~GeV$^2$) is rather uncertain due to the nearby charmed
resonances. The predictions are relatively cleaner in the low-$q^2$
($1 \,{\rm GeV^2} \le q^2 \le 6\, {\rm GeV^2}$) and the high-$q^2$
($14.4\, {\rm GeV^2} \le q^2 \le m_b^2$) regions. Hence we consider
both low-$q^2$ and high-$q^2$  regions in the fit. The branching ratios are 
\cite{Aubert:2004it, Iwasaki:2005sy}
\bea
Br(B \to X_s\, l^+\, l^-)_{{\rm low }\,q^2} &=& (1.60 \pm 0.50)\times 10^{-6}\;,\\
Br(B \to X_s\, l^+\, l^-)_{{\rm high }\,q^2} &=& (0.44 \pm 0.12)\times 10^{-6}\;.
\eea
Both of these branching ratios are sensitive to $\V_{tb}^* \V_{ts}$
and $\V_{t'b}^* \V_{t's}$, i.e. to $A^2$ and $q r e^{i \delta_{cb'}}$,
respectively, to leading order in $\lambda$.

\subsection{\boldmath The $D$ system}

In principle, there can be constraints from $D^0$-$\bar{D}^0$
mixing. In the SM, this mixing arises due to $d$, $s$ and $b$ quarks
in the box diagram. The $b$ contribution is enhanced by a factor of
$(m^2_b-m^2_{s,d})/(m^2_s-m^2_d)$.  On the other hand, it suffers a
strong CKM suppression by a factor of
$|V_{ub}V^*_{cb}|^2/|V_{us}V^*_{cs}|^2$ which is $\sim \lambda^8$.
Thus, $D^0$-$\bar{D}^0$ mixing is dominated by the $d$- and $s$-quark
contributions. As a result, the mixing is small within the SM and
hence sensitive to new physics.

There have been attempts to constrain the CKM4 parameters using
$D^0$-$\bar{D}^0$ mixing (for example, see
Ref.~\cite{Golowich:2007ka}). However, precisely because the $d$ and
$s$ quarks dominate, there can be large long-distance (LD)
contributions to the mixing.  At present, there is no definitive
estimate of these LD effects. Because of this, we do not 
have an accurate enough prediction for $D^0$-$\bar{D}^0$ mixing, 
and this measurement cannot be incorporated in the fit at present.

\section{Results of the Fit}
\label{results}

We perform the fit to 9 CKM4 parameters, using the observables
described in the previous section. We define 
\begin{eqnarray}
\chi^2_{\rm total} &=& \chi^2_{\rm CKM} + \chi^2_{\rm UC} + \chi^2_{ Zbb} + \chi^2_{ ZAb} + \chi^2_{\rm |\epsilon_K|}
+ \chi^2_{K^+\to \pi^+\nu \bar{\nu}} + \chi^2_{\rm mixing}
\nonumber\\&&
+ \chi^2_{\sin 2\beta} + \chi^2_{\gamma}  + \chi^2_{B \to X_s \gamma} + \chi^2_{\rm incl{\hbox{-}}low} 
+ \chi^2_{\rm incl{\hbox{-}}high}\;,
\end{eqnarray}
where the exact definition of each $\chi^2$ contribution is given in
the Appendix~\ref{app-A}.  We perform the fit at two values of $t'$
mass: $m_{t'}=400$ GeV and $m_{t'}=600$ GeV. In addition, we also
perform a fit for the 4 parameters of the CKM matrix in the SM, in
order to check for consistency with the standard fit.  The results of
these fits are summarized in Table~\ref{table:parameters}.  It may be
observed that the $\chi^2$ per degree of freedom is small in each
case, indicating that all the fits are good.  The goodness of fit does
not seem to depend much on the masses of the heavy quarks.

\begin{table}
\begin{tabular}{|c|c|c|c|}
\hline
Parameter & SM & $m_{t'}=400$ GeV & $m_{t'}=600$ GeV \\
\hline
$\lambda$ &$0.227 \pm 0.001$ & $0.227 \pm 0.001$ & $0.227 \pm 0.001$ \\
$A$ &$0.808 \pm 0.021$ & $0.801 \pm 0.022$  &$0.801 \pm 0.022$  \\
$C$ &$0.38 \pm  0.01$ &  $0.42 \pm  0.04$ &$0.42 \pm  0.04$ \\
$\delta_{ub}$ &$1.16  \pm 0.06$ & $1.24  \pm 0.23$ &$1.22  \pm 0.24$ \\
\hline
$p$ & -- & $1.45 \pm  1.20$ & $1.35 \pm  1.56$ \\
$q$ & -- & $0.16 \pm  0.12$ & $0.12 \pm  0.07$\\
$r$ &  -- & $0.30 \pm  0.37$ & $0.19 \pm  0.27$\\
$\delta_{ub'}$& -- & $1.21 \pm 1.59$  & $1.32 \pm 1.76$\\ 
$\delta_{cb'}$ & -- &  $1.10 \pm  1.64$ & $1.25 \pm 1.81$\\
\hline
$\chi^2/d.o.f.$ & $6.64 / 14$ & $6.01 / 11$ & $6.06 / 11$ \\
\hline
\end{tabular}
\caption{The results of the fit to the parameters of CKM and CKM4.
\label{table:parameters}}
\end{table}

The fit for the SM is consistent with that obtained in
Ref.~\cite{pdg}.  As far as the parameters of the three-generation CKM
matrix are concerned, their best-fit values are not affected much by
the addition of a fourth generation. However, the allowed parameter
space for $C$ and $\delta_{ub}$ expands by almost a factor of four.
This is expected, since the constraint on $|\V_{ub}|$ from the $3
\times 3$ unitarity is now relaxed.

On the other hand, the new real parameters $p$, $q$, $r$ are
consistent with zero, which is not surprising since the SM fit is a
good one.  This also is consistent with the observation that no
meaningful constraints are obtained on the new phases $\delta_{ub'}$
and $\delta_{cb'}$: since vanishing $p$, $q$ imply vanishing
$\V_{ub'}, \V_{cb'}$, respectively, the phases of these two CKM4
elements have no significance.

For $m_{t'}=400$ GeV, the maximum values of the parameters $(p,q,r)$ 
are $(2.65,0.28,0.67)$ to $1\sigma$. 
For $m_{t'}=600$ GeV, the $1\sigma$ upper bounds are
$(2.91,0.19,0.46)$.
This indicates that these quantities 
are indeed ${\cal O}(1)$ or smaller, so that the expansion in $\lambda$ 
in the DK parametrization is justified.

\begin{table}[t]
\begin{tabular}{|c|c|c|c|}
\hline
Magnitude & SM & $m_{t'}=400$ GeV & $m_{t'}=600$ GeV \\
\hline
$|\V_{ud}|$ &$0.9743 \pm 0.0002$ & $0.9743 \pm 0.0002$ &$0.9743 \pm 0.0002$  \\
$|\V_{us}|$ &$0.227 \pm 0.001$ &  $0.227 \pm 0.001$ &$0.227 \pm 0.001$ \\
$|\V_{ub}|$ &  $(3.55\pm 0.17) \times 10^{-3}$&  $(3.90\pm 0.38) \times 10^{-3}$ &$(3.91\pm 0.39) \times 10^{-3}$ \\
$|\V_{ub'}|$ & -- & $0.017\pm 0.014$ & $0.016\pm 0.018$\\
\hline
$|\V_{cd}|$ &$0.227 \pm 0.001$ & $0.227 \pm 0.001$ &$0.227 \pm 0.001$ \\
$|\V_{cs}|$ &$0.9743 \pm 0.0002$ &  $0.9743 \pm 0.0002$ &$0.9743 \pm 0.0002$ \\
$|\V_{cb}|$ &$0.042 \pm 0.001$ & $0.041 \pm 0.001$ & $0.041 \pm 0.001$\\
$|\V_{cb'}|$ & --  & $(8.4\pm 6.2) \times 10^{-3}$ &$(6.0\pm 3.8) \times 10^{-3}$ \\
\hline
$|\V_{td}|$ &$0.0086\pm 0.0003$ & $0.009\pm 0.002$ &$0.009\pm 0.001$ \\
$|\V_{ts}|$ &$0.041\pm 0.001$ & $0.041\pm 0.001$ & $0.040\pm 0.001$\\
$|\V_{tb}|$ &1 & $0.998\pm 0.006$ & $0.999\pm 0.003$\\
$|\V_{tb'}|$ & -- &  $0.07\pm 0.08$ &$0.04\pm 0.06$ \\
\hline
$|\V_{t'd}|$ & -- & $0.01\pm 0.01$ & $0.01\pm 0.02$ \\
$|\V_{t's}|$ & -- & $0.01\pm 0.01$ & $0.004\pm 0.010$\\
$|\V_{t'b}|$ & --  & $0.07 \pm 0.08$ &$0.04\pm 0.06$ \\
$|\V_{t'b'}|$ & --  &  $0.998\pm 0.006$ &$0.999\pm 0.003$ \\
\hline
\end{tabular}
\caption{Magnitudes of the CKM4 elements obtained from the fit.
\label{table:magnitudes}}
\end{table}

The magnitudes of CKM4 elements are of special interest, since the
off-diagonal elements are indicative of the mixing between
generations. Table~\ref{table:magnitudes} gives the allowed ranges for
the magnitudes of CKM4 elements, obtained using the fit results in
Table~\ref{table:parameters}.  Clearly the extension to four
generations only expands the allowed ranges of the CKM parameters,
while the allowed values of all of the new parameters of CKM4 (except
$\V_{t'b'}$) are consistent with zero.

\begin{table}[t]
\begin{tabular}{|c|c|c|c|}
\hline
Quantity & SM & $m_{t'}=400$ GeV & $m_{t'}=600$ GeV \\
\hline
$|\V_{tb}^* \V_{td}|$ & $0.0086\pm 0.0003$ & $0.009\pm 0.002$ & $0.009\pm 0.001$ \\
${\rm Arg}(\V_{tb}^* \V_{td})$ & $(-21.5 \pm 1.0)^\circ$ & $(-30.4 \pm 10.3)^\circ$ & $(-27.9 \pm 8.0)^\circ$ \\
\hline
$|\V_{tb}^* \V_{ts}|$ & $0.041\pm 0.001$ & $0.040\pm 0.001$  & $0.040\pm 0.001$  \\
${\rm Arg}(\V_{tb}^* \V_{ts})$ & $(-178.86 \pm 0.06)^\circ$ & $(-178.12 \pm 1.14)^\circ$  & $( -178.12\pm 0.57 )^\circ$  \\
\hline
$|\V_{t'b}^* \V_{t'd}|$ & -- &  $0.0010\pm 0.0015$ & $0.0006\pm 0.0011$ \\
${\rm Arg}(\V_{t'b}^* \V_{t'd})$ & -- &  $( -107.1\pm 106.5)^\circ$ & $( -102.5\pm 112.8)^\circ$ \\
\hline
$|\V_{t'b}^* \V_{t's}|$ & -- & $0.0005 \pm 0.0010$ & $0.0002 \pm 0.0005$ \\
${\rm Arg}(\V_{t'b}^* \V_{t's})$ & -- & $( 37.8\pm 120.3)^\circ$ & $( 40.1\pm 174.1)^\circ$ \\
\hline
\end{tabular}
\caption{Combinations of CKM4 elements that control mixing in the
  $B_d$ and $B_s$ sectors.
\label{table:combinations}}
\end{table}

The combinations of CKM4 matrix elements that control
$B_d$-$\bar{B}_d$ and $B_s$-$\bar{B}_s$ mixing are $\V_{tb}^*
\V_{td}$, $\V_{tb}^* \V_{ts}$, $\V_{t'b} \V_{t'd}$ and $\V_{t'b}
\V_{t's}$. The allowed ranges of these quantities are given in
Table~\ref{table:combinations}.  It may be observed that here the
fourth generation can have maximal impact.  While $|\V_{tb}^*
\V_{ts}|$ is little affected, the allowed range of $|\V_{tb}^*
\V_{td}|$ is increased by up to a factor of 6-7.  Moreover, the
allowed range of the phase of $\V_{tb}^* \V_{td}$ is expanded by $\sim
10$, while that of the phase of $\V_{tb}^* \V_{ts}$ is larger by a
factor of $\sim 20$ at $m_{t'} = 400$ GeV.  Since the phases of
$\V_{t'b}^* \V_{t'd}$ and $\V_{t'b}^* \V_{t's}$ are essentially
unconstrained, they can influence $B_d$ and $B_s$ mixing to a large
extent. In particular, the $B_s$-$\bar{B}_s$ mixing phase can be very
large, as suggested by the recent measurements from $B_s \to J/\psi
\phi$ decays \cite{jpsiphi}. The combinations  $\V_{ts}^* \V_{td}$ and 
$\V_{t's}^* \V_{t'd}$ also contribute to rare $K$ decays, 
and hence significant deviations of these quantities 
from the SM can also leave their imprints in the rare $K$ decays.

\section{Discussion}
\label{concl}

In this paper we consider the extension of the standard model (SM) to
four generations. Using input from many flavor-physics processes, we
perform a $\chi^2$ fit to constrain the elements of the $4\times 4$
CKM quark-mixing matrix (CKM4). The fit takes into account both
experimental errors and theoretical uncertainties. Although we do not
include the oblique parameters in our fit, we do take values for the
masses of the fourth-generation quarks that are consistent with the
oblique corrections.

At this stage, several comments are in order. 
\begin{itemize}

\item The best-fit values of all three new real parameters of the CKM4
  matrix are consistent with zero. Since the fit to the SM is also
  excellent -- $\chi^2/d.o.f.\ = 6.64/14$, corresponding to a
  goodness-of-fit of $92\%$ -- we must conclude that the addition of a
  fourth generation is not necessary to get a better fit to the data.

\item We find $\V_{tb} = 0.998\pm 0.006 $ for $m_{t'}=400$ GeV and
  $\V_{tb} = 0.999\pm 0.003 $ for $m_{t'}=600$ GeV.  Thus, at
  $3\sigma$, we have $\V_{tb} \ge 0.98$.  Therefore the SM4 cannot account for any
  large deviation of $V_{tb}$ from unity.

\item In many previous analyses, it is mentioned that any mixing
  between the third and fourth generations is small. We find that this
  is indeed the case -- the results of the fit constrain the matrix
  elements describing the mixing of the ordinary and fourth-generation
  quarks to be $|\V_{ub'}|< 0.06$, $|\V_{cb'}|< 0.027$, and
  $|\V_{tb'}|< 0.31$ at $3\sigma$.

\item However, the allowed parameter ranges still allow large
  deviations from the SM as far as the magnitudes and phases of the
  quantities $\V_{tb}^* \V_{td}$ and $\V_{tb}^* \V_{ts}$ are
  concerned. With additional new-physics contributions involving
  $\V_{t'b} \V_{t'd}$ and $\V_{t'b} \V_{t's}$, it may be possible to
  get significant new-physics signals in $B_d$ and/or $B_s$ mixing,
  which could be the most incisive probes of the fourth generation.

\item The value of $|V_{ub}|$ required to explain the recent 
    measurement of  $Br(B^+ \to \tau^+ \nu_\tau)$  
    is $2.8\sigma$ larger than that obtained from the global fit to $|V_{ub}|$ 
    otherwise, within the SM \cite{ckmfitter:ichep10}. 
    Our fit indicates that the best fit for 
    $|\V_{ub}|$ shifts to higher values with SM4, and the error on 
    this quantity also increases. As a result, the SM4 may be able 
    to account for this measurement much better than the SM.

\end{itemize}


\section*{Acknowledgments}

We thank Michael Chanowitz, Paul Langacker,
Alexey Petrov, Seungwon Baek, S. Uma Sankar and Georges Azuelos for
useful discussions and communications.  This work was financially
supported by NSERC of Canada (AKA, DL).


\appendix


\section{\boldmath The $\chi^2$ function}
\label{app-A}

We define our $\chi^2$ function to be
\begin{eqnarray}
\chi^2_{\rm total} &=& \chi^2_{\rm CKM} + \chi^2_{\rm UC} + \chi^2_{ Zbb} 
+ \chi^2_{ ZAb} + \chi^2_{\rm |\epsilon_K|}
+ \chi^2_{K^+\to \pi^+\nu \bar{\nu}} + \chi^2_{\rm mixing} 
\nonumber\\
&& + \chi^2_{\sin 2\beta} + \chi^2_{\gamma}  + \chi^2_{B \to X_s \gamma} 
+ \chi^2_{\rm incl{\hbox{-}}low} + \chi^2_{\rm incl{\hbox{-}}high}\;.
\end{eqnarray}
The components of this function are defined below.
\begin{itemize}

\item For the direct measurements of the magnitudes of the elements, 
\bea
\chi^2_{\rm CKM} &=& \Big( \frac{|\V_{us}|-0.2255}{0.0019} \Big)^2 
+ \Big( \frac{|\V_{ud}|-0.97418}{0.00027} \Big)^2
+ \Big( \frac{|\V_{cs}|-1.04}{0.06} \Big)^2 
\nonumber\\&& \hskip-1truecm
 +~\Big( \frac{|\V_{cd}|-0.230}{0.011} \Big)^2
+ \Big( \frac{|\V_{ub}|-0.00393}{0.00036} \Big)^2 
+ \Big( \frac{|\V_{cb}|-0.0412}{0.0011} \Big)^2\;.
\eea
\bea
\chi^2_{\rm UC} &=& \Big( \frac{|\V_{ub'}|^2 - 0.0001}{0.0011} \Big)^2 
+ \Big( \frac{|\V_{cb'}|^2 +  0.136}{0.125} \Big)^2
\nonumber\\
&& \hskip-1truecm
+~\Big( \frac{(|\V_{td}|^2 + |\V_{t'd}|^2) + 0.002}{0.005} \Big)^2
+ \Big( \frac{(|\V_{ts}|^2 + |\V_{t's}|^2) + 0.134}{0.125} \Big)^2\;.
\eea

\item For the $Z \to b \bar{b}$ decay,
\beq
\chi^2_{Zbb} = \Big( \frac{R_{bb}-0.216}{0.001} \Big)^2\;, 
\eeq
where $R_{bb}$ is defined in Eq.~(\ref{rbbf}) and 
\beq
\chi^2_{ZAb} = \Big( \frac{A_{b}-0.923}{0.020} \Big)^2\;, 
\eeq
where $A_{b}$ is defined in Eq.~(\ref{zafb}).

\item For $K$ mixing,
\beq
\chi^2_{\rm |\epsilon_K|} = \Big( \frac{|\epsilon_K| - 0.00232}{0.00046} 
\Big)^2\;, 
\eeq
where $\epsilon_K$ is defined in Eq.~(\ref{cpvkpipi}). 
Here experimental and theoretical errors are added in quadrature, 

\item Next, we have
\beq
\chi^2_{K^+\to \pi^+\nu \bar{\nu}} =\Big( \frac{Br(K^+\to \pi^+\nu\bar{\nu}) - 1.7\times 10^{-10}}{1.1\times 10^{-10}} \Big)^2\;,
\eeq
with $Br(K^+\to \pi^+\nu\bar{\nu})$ as in Eq.~(\ref{brkpinunu}).

\item In $B$-meson mixing, 
\beq
\chi^2_{\rm mixing} = \Big( \frac{M_R-22.20}{1.04} \Big)^2\;,
\eeq
with $M_R$ as defined in Eq.~(\ref{eq:mr}) and
\beq
\frac{\Delta M_s}{\Delta M_d}\,\frac{m_{B_d}}{m_{B_s}}\,\frac{1}{\xi^2} = 22.40 \pm 1.04 \;.
\eeq

\item For CP violation in $B_d \to J/\psi K_S$,
\beq
\chi^2_{\sin 2\beta} = \Big( \frac{S_{J/\psi K_S}-0.672}{0.024} \Big)^2\;,
\eeq
with $S_{J/\psi K_S}$ defined as in Eq.~(\ref{eq:sin2b}).

\item For the CKM angle $\gamma$
\beq
\chi^2_{\gamma} = \Big( \frac{\delta_{ub}- 75~(\pi/180)}{22~(\pi/180)} \Big)^2\;,
\eeq

\item For the radiative decay
\beq
\chi^2_{B \to X_s \gamma} =\Big( \frac{100\,R -0.330} {0.041}\Big)^2\;,
\eeq
with $R$ as defined in Eq.~(\ref{R}). Here experimental and theoretical 
errors are added in quadrature. 

\item For the leptonic decay,
\beq
\chi^2_{\rm incl{\hbox{-}}low} = \Big( \frac{Br(B \to X_s\, l^+\, l^-)_{{\rm low }\,q^2}\times 10^{6} -1.6} {0.55}\Big)^2\;.
\eeq
$Br(B \to X_s\, l^+\, l^-)_{{\rm low }\,q^2}$ has been obtained 
by integrating Eq.~(\ref{eq:brincl}) within the limits 
($1 \,{\rm GeV^2} \le q^2 \le 6\, {\rm GeV^2}$). 
Here experimental and theoretical errors are added in quadrature. 

\item Similarly,
\beq
\chi^2_{\rm incl{\hbox{-}}high} = \Big( \frac{Br(B \to X_s\, l^+\, l^-)_{{\rm high }\,q^2}\times 10^{6} -0.44} {0.14}\Big)^2\;,
\eeq
where $Br(B \to X_s\, l^+\, l^-)_{{\rm high }\,q^2}$ has been obtained 
by integrating Eq.~(\ref{eq:brincl}) within the limits 
($14.4\, {\rm GeV^2} \le q^2 \le m_b^2$). 

\end{itemize}


\begin{thebibliography}{10}

\bibitem{Holdom:2009rf}
  B.~Holdom, W.~S.~Hou, T.~Hurth, M.~L.~Mangano, S.~Sultansoy and G.~Unel,
  PMC Phys.\  A {\bf 3}, 4 (2009)
  [arXiv:0904.4698 [hep-ph]].

\bibitem{Hou:2005yb}
  W.~S.~Hou, M.~Nagashima and A.~Soddu,
  Phys.\ Rev.\  D {\bf 72}, 115007 (2005)
  [arXiv:hep-ph/0508237].

\bibitem{Hou:2006mx}
  W.~S.~Hou, M.~Nagashima and A.~Soddu,
  Phys.\ Rev.\  D {\bf 76}, 016004 (2007)
  [arXiv:hep-ph/0610385].

\bibitem{Soni:2008bc}
  A.~Soni, A.~K.~Alok, A.~Giri, R.~Mohanta and S.~Nandi,
  Phys.\ Lett.\  B {\bf 683}, 302 (2010)
  [arXiv:0807.1971 [hep-ph]].

\bibitem{Soni:2010xh}
  A.~Soni, A.~K.~Alok, A.~Giri, R.~Mohanta and S.~Nandi,
  Phys.\ Rev.\  D {\bf 82}, 033009 (2010)
  [arXiv:1002.0595 [hep-ph]].

\bibitem{Buras:2010pi}
  A.~J.~Buras, B.~Duling, T.~Feldmann, T.~Heidsieck, C.~Promberger and S.~Recksiegel,
  arXiv:1002.2126 [hep-ph].

\bibitem{Hou:2010mm}
  W.~S.~Hou and C.~Y.~Ma,
  Phys.\ Rev.\  D {\bf 82}, 036002 (2010)
  [arXiv:1004.2186 [hep-ph]].

\bibitem{EW-breaking}
  C.~T.~Hill, M.~A.~Luty and E.~A.~Paschos,
  Phys.\ Rev.\  D {\bf 43}, 3011 (1991);
  B.~Holdom,
  Phys.\ Rev.\ Lett.\  {\bf 57}, 2496 (1986)
  [Erratum-ibid.\  {\bf 58}, 177 (1987)];
  T.~Elliott and S.~F.~King,
  Phys.\ Lett.\  B {\bf 283}, 371 (1992);
  G.~Burdman and L.~Da Rold,
  JHEP {\bf 0712}, 086 (2007)
  [arXiv:0710.0623 [hep-ph]];
  W.~A.~Bardeen, C.~T.~Hill and M.~Lindner,
  Phys.\ Rev.\  D {\bf 41}, 1647 (1990).

\bibitem{EW-transition}
  S.~W.~Ham, S.~K.~Oh and D.~Son,
  Phys.\ Rev.\  D {\bf 71}, 015001 (2005)
  [arXiv:hep-ph/0411012];
  R.~Fok and G.~D.~Kribs,
  Phys.\ Rev.\  D {\bf 78}, 075023 (2008)
  [arXiv:0803.4207 [hep-ph]];
Y.~Kikukawa, M.~Kohda and J.~Yasuda,
  Prog.\ Theor.\ Phys.\  {\bf 122}, 401 (2009)
  [arXiv:0901.1962 [hep-ph]].
  M.~S.~Carena, A.~Megevand, M.~Quiros and C.~E.~M.~Wagner,
  Nucl.\ Phys.\  B {\bf 716}, 319 (2005)
  [arXiv:hep-ph/0410352];
    W.~S.~Hou,
  Chin.\ J.\ Phys.\  {\bf 47}, 134 (2009)
  [arXiv:0803.1234 [hep-ph]].

\bibitem{EW-constraint}
  M.~Maltoni, V.~A.~Novikov, L.~B.~Okun, A.~N.~Rozanov and M.~I.~Vysotsky,
  Phys.\ Lett.\  B {\bf 476}, 107 (2000)
  [arXiv:hep-ph/9911535];
  V.~A.~Novikov, L.~B.~Okun, A.~N.~Rozanov, M.~I.~Vysotsky and V.~P.~Yurov,
  Mod.\ Phys.\ Lett.\  A {\bf 10}, 1915 (1995)
  [Erratum-ibid.\  A {\bf 11}, 687 (1996)];
  N.~J.~Evans,
  Phys.\ Lett.\  B {\bf 340}, 81 (1994)
  [arXiv:hep-ph/9408308];
  H.~J.~He, N.~Polonsky and S.~f.~Su,
  Phys.\ Rev.\  D {\bf 64}, 053004 (2001)
  [arXiv:hep-ph/0102144];
  V.~A.~Novikov, L.~B.~Okun, A.~N.~Rozanov and M.~I.~Vysotsky,
  Phys.\ Lett.\  B {\bf 529}, 111 (2002)
  [arXiv:hep-ph/0111028].

 \bibitem{Kribs:2007nz}
 G.~D.~Kribs, T.~Plehn, M.~Spannowsky and T.~M.~P.~Tait,
 Phys.\ Rev.\  D {\bf 76}, 075016 (2007)
 [arXiv:0706.3718 [hep-ph]].

\bibitem{erler-langacker}
  J.~Erler and P.~Langacker,
  Phys.\ Rev.\ Lett.\  {\bf 105}, 031801 (2010)
  [arXiv:1003.3211 [hep-ph]].


\bibitem{Chanowitz:1978mv}
  M.~S.~Chanowitz, M.~A.~Furman and I.~Hinchliffe,
  Nucl.\ Phys.\  B {\bf 153}, 402 (1979).

\bibitem{Marciano:1989ns}
  W.~J.~Marciano, G.~Valencia and S.~Willenbrock,
  Phys.\ Rev.\  D {\bf 40}, 1725 (1989).

\bibitem{Abazov:2009ii}
  V.~M.~Abazov {\it et al.}  [D0 Collaboration],
  Phys.\ Rev.\ Lett.\  {\bf 103}, 092001 (2009)
  [arXiv:0903.0850 [hep-ex]].

\bibitem{Aaltonen:2009jj}
  T.~Aaltonen {\it et al.}  [CDF Collaboration],
  Phys.\ Rev.\ Lett.\  {\bf 103}, 092002 (2009)
  [arXiv:0903.0885 [hep-ex]].

\bibitem{Group:2009qk}
  T.~E.~W.~Group  [CDF and D0 Collaboration],
  arXiv:0908.2171 [hep-ex].

\bibitem{pdg} K.~Nakamura {\it et al.}  [Particle Data Group],
  J.\ Phys.\ G {\bf 37}, 075021 (2010).
 
 \bibitem{Alwall:2006bx}
  J.~Alwall {\it et al.},
  Eur.\ Phys.\ J.\  C {\bf 49}, 791 (2007)
  [arXiv:hep-ph/0607115].

\bibitem{Herrera:2008yf}
  J.~A.~Herrera, R.~H.~Benavides and W.~A.~Ponce,
  Phys.\ Rev.\  D {\bf 78}, 073008 (2008)
  [arXiv:0810.3871 [hep-ph]].

 \bibitem{Bobrowski:2009ng}
M.~Bobrowski, A.~Lenz, J.~Riedl and J.~Rohrwild,
  Phys.\ Rev.\  D {\bf 79}, 113006 (2009)
  [arXiv:0902.4883 [hep-ph]].
  
\bibitem{Chanowitz:2009mz}
  M.~S.~Chanowitz,
  Phys.\ Rev.\  D {\bf 79}, 113008 (2009)
  [arXiv:0904.3570 [hep-ph]].

\bibitem{Eberhardt:2010bm}
  O.~Eberhardt, A.~Lenz and J.~Rohrwild,
  arXiv:1005.3505 [hep-ph].

\bibitem{Nandi:2010zx}
  S.~Nandi and A.~Soni,
  arXiv:1011.6091 [hep-ph].

\bibitem{Kim:2007zzg}
  C.~S.~Kim and A.~S.~Dighe,
  Int.\ J.\ Mod.\ Phys.\  E {\bf 16}, 1445 (2007)
  [arXiv:0710.1681 [hep-ph]].
  
\bibitem{Alok:2008dj} 
  A.~K.~Alok, A.~Dighe and S.~Ray,
  Phys.\ Rev.\  D {\bf 79}, 034017 (2009)
  [arXiv:0811.1186 [hep-ph]].

\bibitem{Wolfenstein} L.~Wolfenstein,
  Phys.\ Rev.\ Lett.\  {\bf 51}, 1945 (1983).

\bibitem{minuit}
  F.~James and M.~Roos,
  Comput.\ Phys.\ Commun.\  {\bf 10}, 343 (1975).

\bibitem{Bernabeu}
 J.~Bernabeu, A.~Pich and A.~Santamaria,
  Nucl.\ Phys.\  B {\bf 363}, 326 (1991).

\bibitem{Chetyrkin:1990kr}
  K.~G.~Chetyrkin and J.~H.~Kuhn,
  Phys.\ Lett.\  B {\bf 248}, 359 (1990).

\bibitem{Kniehl:1989bb}
  B.~A.~Kniehl and J.~H.~Kuhn,
  Phys.\ Lett.\  B {\bf 224}, 229 (1989).

\bibitem{Yanir:2002cq}
  T.~Yanir,
  JHEP {\bf 0206}, 044 (2002)
  [arXiv:hep-ph/0205073].

\bibitem{lepewwg}
The ALEPH, CDF, D0, DELPHI, L3, OPAL, SLD Collaborations, 
the LEP Electroweak Working Group, the Tevatron Electroweak 
Working Group, and the SLD Electroweak and heavy flavour groups,  
arXiv:0811.4682 [hep-ex]. 

\bibitem{vysotsky}
  M.~I.~Vysotsky, V.~A.~Novikov, L.~B.~Okun and A.~N.~Rozanov,
  Phys.\ Usp.\  {\bf 39}, 503 (1996)
  [Usp.\ Fiz.\ Nauk {\bf 166}, 539 (1996)]
  [arXiv:hep-ph/9606253].


\bibitem{Akhundov:1985fc}
  A.~A.~Akhundov, D.~Y.~Bardin and T.~Riemann,
  Nucl.\ Phys.\  B {\bf 276}, 1 (1986).

\bibitem{Beenakker:1988pv}
  W.~Beenakker and W.~Hollik,
  Z.\ Phys.\  C {\bf 40}, 141 (1988).

\bibitem{Bernabeu:1987me}
  J.~Bernabeu, A.~Pich and A.~Santamaria,
  Phys.\ Lett.\  B {\bf 200}, 569 (1988).

\bibitem{Lynn:1990hd}
  B.~W.~Lynn and R.~G.~Stuart,
  Phys.\ Lett.\  B {\bf 252}, 676 (1990).


\bibitem{Buras:2008nn}
  A.~J.~Buras and D.~Guadagnoli,
  Phys.\ Rev.\  D {\bf 78}, 033005 (2008)
  [arXiv:0805.3887 [hep-ph]].

\bibitem{Buras:1997fb}
  A.~J.~Buras and R.~Fleischer,
  Adv.\ Ser.\ Direct.\ High Energy Phys.\  {\bf 15}, 65 (1998)
  [arXiv:hep-ph/9704376].
  
\bibitem{Laiho:2009eu}
  J.~Laiho, E.~Lunghi and R.~S.~Van de Water,
  Phys.\ Rev.\  D {\bf 81}, 034503 (2010)
  [arXiv:0910.2928 [hep-ph]].

\bibitem{Inami:1980fz}
  T.~Inami and C.~S.~Lim, 
  Prog.\ Theor.\ Phys.\  {\bf 65}, 297 (1981) 
  [Erratum-ibid.\  {\bf 65}, 1772 (1981)]. 
  
\bibitem{Herrlich:1993yv}
  S.~Herrlich and U.~Nierste,
  Nucl.\ Phys.\  B {\bf 419}, 292 (1994)
  [arXiv:hep-ph/9310311].

\bibitem{Herrlich:1995hh}
  S.~Herrlich and U.~Nierste,
  Phys.\ Rev.\  D {\bf 52}, 6505 (1995)
  [arXiv:hep-ph/9507262].

\bibitem{Herrlich:1996vf}
  S.~Herrlich and U.~Nierste,
  Nucl.\ Phys.\  B {\bf 476}, 27 (1996)
  [arXiv:hep-ph/9604330].

\bibitem{Buras:1990fn}
  A.~J.~Buras, M.~Jamin and P.~H.~Weisz,
  Nucl.\ Phys.\  B {\bf 347}, 491 (1990).

  \bibitem{Hattori:1999ap}
  T.~Hattori, T.~Hasuike and S.~Wakaizumi,
  Phys.\ Rev.\  D {\bf 60}, 113008 (1999)
  [arXiv:hep-ph/9908447].

\bibitem{Buchalla:1995vs}
  G.~Buchalla, A.~J.~Buras and M.~E.~Lautenbacher,
  Rev.\ Mod.\ Phys.\  {\bf 68}, 1125 (1996)
  [arXiv:hep-ph/9512380].

\bibitem{Rein:1989tr}
  D.~Rein and L.~M.~Sehgal,
  Phys.\ Rev.\  D {\bf 39}, 3325 (1989).

\bibitem{Hagelin:1989wt}
  J.~S.~Hagelin and L.~S.~Littenberg,
  Prog.\ Part.\ Nucl.\ Phys.\  {\bf 23}, 1 (1989).

\bibitem{Buchalla:1993wq}
  G.~Buchalla and A.~J.~Buras,
  Nucl.\ Phys.\  B {\bf 412}, 106 (1994)
  [arXiv:hep-ph/9308272].
  
\bibitem{Mescia:2007kn}
  F.~Mescia and C.~Smith,
  Phys.\ Rev.\  D {\bf 76}, 034017 (2007)
  [arXiv:0705.2025 [hep-ph]].
  
\bibitem{CDFD0} S.~Malde  [CDF Collaboration and D0 Collaboration],
  arXiv:0909.5644 [hep-ex].

\bibitem{BsNPdecay} C.~W.~Chiang, A.~Datta, M.~Duraisamy, D.~London, M.~Nagashima and A.~Szynkman,
  JHEP {\bf 1004}, 031 (2010)
  [arXiv:0910.2929 [hep-ph]].

\bibitem{hfag}
  E.~Barberio {\it et al.}  [Heavy Flavor Averaging Group],
  arXiv:0808.1297 [hep-ex].

\bibitem{misiak}
  M.~Misiak and M.~Steinhauser,
  Nucl.\ Phys.\  B {\bf 764}, 62 (2007)
  [arXiv:hep-ph/0609241];
  M.~Misiak {\it et al.},
  Phys.\ Rev.\ Lett.\  {\bf 98}, 022002 (2007)
  [arXiv:hep-ph/0609232].

\bibitem{Buras:1994dj}
  A.~J.~Buras and M.~Munz,
  Phys.\ Rev.\  D {\bf 52}, 186 (1995)
  [arXiv:hep-ph/9501281].

\bibitem{Nir:1989rm}
  Y.~Nir,
  Phys.\ Lett.\  B {\bf 221}, 184 (1989).

\bibitem{Aubert:2004aw}
  B.~Aubert {\it et al.}  [BABAR Collaboration],
  Phys.\ Rev.\ Lett.\  {\bf 93}, 011803 (2004)
  [arXiv:hep-ex/0404017].

\bibitem{Aubert:2004it}
  B.~Aubert {\it et al.}  [BABAR Collaboration],
  Phys.\ Rev.\ Lett.\  {\bf 93}, 081802 (2004)
  [arXiv:hep-ex/0404006].

\bibitem{Iwasaki:2005sy}
  M.~Iwasaki {\it et al.}  [Belle Collaboration],
  Phys.\ Rev.\  D {\bf 72}, 092005 (2005)
  [arXiv:hep-ex/0503044].
  
  \bibitem{Golowich:2007ka}
  E.~Golowich, J.~Hewett, S.~Pakvasa and A.~A.~Petrov,
  Phys.\ Rev.\  D {\bf 76}, 095009 (2007)
  [arXiv:0705.3650 [hep-ph]].

\bibitem{jpsiphi}
T. Aaltonen et al. (CDF Collaboration), CDF Note
No. CDF/PHYS/BOTTOM/CDFR/9787, 2009; V. M.
Abazov et al. (D0 Collaboration), D0 Note No. 5928-
CONF, 2009.

\bibitem{ckmfitter:ichep10} S. T'Jampens [CKMFitter collaboration], Talk at ICHEP 2010.
http://indico.cern.ch/getFile.py/access?contribId=190\&sessionId=53\&resId=0\& \\materialId=slides\&confId=73513

\end{thebibliography}
\end{document}